\documentstyle{europhys}

\def\And{{\rm and\ }}

\def\stars{\bigskip\centerline{***}\medskip}

\newif\ifboo \boofalse

\def\Review#1{\boofalse{\it #1},}
\def\Name#1{{\sc #1},}
\def\Vol#1{\ifboo Vol. {\bf #1}\else{\bf #1}\fi}
\def\Year#1{\ifboo #1\else(#1)\fi}
\def\Book#1{\bootrue{\it #1},}
\def\Page#1{\ifboo {\rm p. #1}\else{\rm #1}\fi}

\begin{document}
\euro{}{}{}{}
\Date{}
\title{Intrinsically localized chaos in discrete nonlinear extended systems}

\author{P. J. Mart\'{\i}nez\inst{1}, L. M. Flor\'{\i}a\inst{2}, 
F. Falo\inst{2} and J. J. Mazo\inst{2,}\inst{3}}
\institute{
    \inst{1} Group of Theory and Simulation of Complex Systems, 
Departamento de F\'{\i}sica Aplicada, CSIC-Universidad de Zaragoza, E-50009 Zaragoza, Spain \\
    \inst{2} Group of Theory and Simulation of Complex Systems, 
Departamento de F\'{\i}sica de la Materia Condensada, CSIC-Universidad de Zaragoza, E-50009 Zaragoza, Spain \\
    \inst{3} Department of Electrical Engineering and Computer Science, 
Massachusetts Institute of Technology, Cambridge, MA 02319, USA}
\rec{}{}
\pacs{
\Pacs{05}{45$+$b}{Theory and models of chaotic systems}
\Pacs{03}{20$+$i}{Classical mechanics of discrete systems: general
mathematics aspects}
\Pacs{74}{50$+$r}{Proximity effects, weak links, tunneling phenomena and
Josephson  effects}
}
\maketitle

\begin{abstract}
The phenomenon of intrinsic localization in discrete nonlinear extended 
systems, {\em i.e.} the (generic) existence of discrete breathers, is shown 
to be not restricted to periodic solutions but it also extends to more complex 
(chaotic) dynamical behaviour. We illustrate this with two different forced 
and damped systems exhibiting this type of solutions: In an anisotropic 
Josephson junction ladder, we obtain intrinsically localized chaotic 
solutions by following periodic rotobreather solutions through a cascade 
of period-doubling bifurcations. In an array of forced and damped van der Pol 
oscillators, they are obtained by numerical continuation (path-following) 
methods from the uncoupled limit, where its existence is trivially 
ascertained, following the ideas of the anticontinuum limit.
\end{abstract}

Discrete homogeneous arrays of (hamiltonian and non-hamiltonian) nonlinear 
oscillators (or rotors) exhibit generic solutions which are time-periodic 
and (typically exponentially) localized in space. These solutions are called 
{\em discrete breathers} by analogy with non-topological localized solutions 
of certain PDE's. In contrast with continuous "bona fide" breathers, discrete 
breathers posses a remarkable structural stability, and thus genericity. This 
localization is often referred to as {\em intrinsic} to stress the fact that 
the system is homogeneous (no impurities or disorder are present). For an 
updated and comprehensive review on discrete breathers, see\cite{FlachWillis}.

A general schematic way to describe a discrete breather in a one-dimensional
lattice is the following: Let us consider the phase space $\Gamma_{s}$ of a
single oscillator (or rotor), so that the phase space $\Gamma$ of the
network is the cartesian product of the single site phase spaces. Let denote
by $A$, $B$ periodic orbits in $\Gamma_s$, eventually projections of
trajectories of $\Gamma$ onto $\Gamma_s$. A discrete breather is a solution 
\begin{equation}
\label{Schembreather}
\{\phi_{i}(t)\} \equiv \{ \ldots,B_{-2}\;,B_{-1}\;,A\;,B_{1}\;,B_{2}\;,\ldots\}
\end{equation}
with $\lim_{|i| \rightarrow \infty} B_{i} = B_{\infty}$, and 
$B_{\infty}\neq A$. Archetypical examples are Klein-Gordon hamiltonian 
breathers, where $A$ is a periodic cycle of frequency $\omega_b$ in the 
$(\phi,\dot{\phi})$ phase space, $B_{\infty}$ is the rest solution $(0,0)$, 
and $B_{i}$ are $\omega_b$-cycles with exponentially decreasing amplitude. In 
the case of forced and damped arrays, $A$ and $B_{\infty}$ are usually 
$\omega_b$-cycles of different amplitude. If $A$ is an $\omega_b$-cycle 
non-homotopic to zero ({\em i. e.} the central oscillator rotates), the term 
{\em rotobreather} is used\cite{TakenoPeyrard,Aubry97}.

In this letter we present numerical evidences as well as plausibility arguments 
strongly supporting the conclusion that the phenomenon of intrinsic 
localization in discrete nonlinear extended systems is not restricted to 
time-periodic solutions, but it extends to more complex (chaotic) behaviour 
in a generic way for damped and forced systems. More specifically, 
we show below examples of solutions of 
the type schematized in (\ref{Schembreather}), where $A$ is a {\em chaotic 
trajectory}, $B_{\infty}$ is a "regular" $\omega_b$-cycle, and $B_{i}$ are 
"noisy" cycles, with "noise intensity" exponentially decreasing to zero as 
$|i|$ grows. The first example concerns the operation of a Josephson junction 
device, the {\em Josephson junction ladder}, which has recently received some 
attention from both theoretical\cite{Mazo,FloriaEuro,Barahona2,Martinez} and 
experimental\cite{Barahona1} sides, in connection to the relevance of 
nonlinear dynamics of discrete systems in Condensed Matter Physics. 
The second example, though also experimentally realizable, serves us 
to illuminate possible pathways towards a rigorous characterization 
of the genericity of intrinsically localized chaos in discrete nonlinear 
extended systems, in the spirit of the ideas of the, so called, 
{\em anticontinuum limit}\cite{Aubry94-95,MackayAubry,MarinAubry} approach 
to intrinsic localization. We end with a short discussion on the implausibility
of existence of this type of solutions as exact ones in hamiltonian 
arrays. Earlier numerical observations of localized chaotic solutions 
seems to have been reported in\cite{Kaneko}(coupled map lattices) 
and\cite{Greenfield}(domain walls in a parametrically excited lattice of 
oscillators). Our results establish a precise (and very general) link between 
situations of spatio-temporal complex behaviour in spatially extended discrete
systems and the emergent new results and powerful methods of intrinsic
localization.

Recent theoretical analyses of the dynamics of an anisotropic
Josephson junction ladder (see figure~\ref{ladder}) with injected ac
currents~\cite{FloriaEuro} have shown the existence of discrete breathers 
as attracting solutions of the equations of motion describing the dynamics 
of the system in the framework of the resistively and capacitively shunted 
junction (RCSJ) approach~\cite{Tinkham}. The existence of discrete breathers 
in Josephson junction arrays should indeed be regarded as generic, given the 
connection between the general description of these systems in terms of the 
superconducting Ginzburg-Landau order parameter 
$\Psi(\vec{x})=|\Psi(\vec{x})|\exp(i\theta(\vec{x}))$, where $\vec{x}$ 
denotes the superconducting island position, and the \emph{discrete nonlinear 
Schr\"odinger equation}, for the case of ideal (perfect insulating)
junctions~\cite{FloriaEuro}. In fact, the quantum Hamiltonian of a single
ideal Josephson junction corresponds to the problem of two coupled
anharmonic quantum oscillators, for which the asymmetric classical
breather solutions have been shown to persist in the quantum regime as
very long lifetime states~\cite{AubryFlach} (see also \cite{Eilbeck}). 
When the energy cost to add an extra Cooper pair on a neutral superconducting 
island (\emph{charging energy} $E_c$) is much lower than the tunneling energy
(\emph{Josephson energy} $E_J$) the superconducting phase
$\theta(\vec{x})$ becomes a good (very weakly fluctuating) variable to
describing the island state, thus validating the RCSJ
approach~\cite{Tinkham}. This is the situation when the
superconducting islands are of macroscopic size. The validity of the
RCSJ approach in the regime $E_c/E_J \ll 1$ is a well established
issue and its predictions fit very well with experiments\cite{Watanabe}.

\begin{figure}
\vbox to 4cm{\vfill\centerline{\fbox{Here is the figure}}\vfill}
\caption{Schematic picture of the JJ ladder showing the injection of the
currents in the array.}
\label{ladder}
\end{figure}

$\theta_i$ and $\theta'_i$ will denote, respectively, the phases of 
upper and lower islands at site $i$ in the ladder; the currents 
$I(t) = I_{ac}\cos(\omega t)$ are injected into the islands in the 
upper row and extracted from those in the lower row; $(J_x,\; \epsilon_x)$ 
are the junction characteristics for junctions in horizontal links and 
$(J_y,\; \epsilon_y)$ for junctions in vertical links. With the change 
of variables $\chi_i=\frac{1}{2}(\theta_i+\theta'_i)$,
$\phi_i=\frac{1}{2}(\theta_i-\theta'_i)$, the RCSJ
equations~\cite{FloriaEuro} are

\label{chiphieqns}
\begin{eqnarray}
  \ddot{\chi}_i = & J_x \left[
    \sin(\chi_{i+1}-\chi_i)\cos(\phi_{i+1}-\phi_i)+ \right.
    \left. \sin(\chi_{i-1}-\chi_i)\cos(\phi_{i-1}-\phi_i)\right] \nonumber \\
   &
+\epsilon_x\left(\dot{\chi}_{i+1}+\dot{\chi}_{i-1}-2\dot{\chi}_i\right)
\label{equationa}
\end{eqnarray}
\begin{eqnarray}
  \ddot{\phi}_i = & J_x \left[
    \cos(\chi_{i+1}-\chi_i)\sin(\phi_{i+1}-\phi_i)+ \right.
   \left. \cos(\chi_{i-1}-\chi_i)\sin(\phi_{i-1}-\phi_i)\right]\nonumber \\
   &
  +\epsilon_x\left(\dot{\phi}_{i+1}+\dot{\phi}_{i-1}-2\dot{\phi}_i\right)
  -J_y \sin(2\phi_i) -2\epsilon_y \dot{\phi}_i - I(t)
\label{equationb}
\end{eqnarray}

With uniform initial conditions in the "center of mass" coordinates
and momenta: $\chi_i$ and $\dot{\chi}_i$ independent of $i$, equations
(\ref{equationa}) have the solution $\chi_i(t) = \Omega t + \alpha$
for all $i$; this effectively decouples equations (\ref{equationb})
for the $\phi_i$ variables from equations (\ref{equationa}) for the
$\chi_i$ variables. Then, using efficient {\em continuation
methods}~\cite{MackayAubry,MarinAubry} from the uncoupled (anticontinuum) 
limit ($J_x=\epsilon_x=0$), one easily computes discrete breather solutions;
these turn out to be attractors of the dynamics of the ladder in a
wide range of parameter values.

We will concentrate on the \emph {rotobreather} type of solutions, in 
which the phase half-difference $\phi_{j^{*}}$ through a vertical junction 
rotates, while the rest $\phi_{i}$ ($ i \neq j^{*}$)
oscillate, and the "center of mass" variables $\chi_i$ remain
uniformly at rest ($\Omega =\alpha=0$; note that any other values for
these parameters, fixed by the uniform initial conditions, would show
the same behavior).  The period of the rotobreather solution is $T_b =
2\pi/\omega_b = 4\pi/\omega$, where $\omega$ is the frequency of the
external currents. 

By performing the Floquet analysis of rotobreather solutions, one can 
determine the regions of linear stability in parameter space, whose 
borders correspond to different types of bifurcations\cite{Martinez}. 
One of them (which occurs typically when varying the external frequency 
$\omega$) is a period-doubling bifurcation: The (destabilizing) 
eigenvector of the Floquet matrix, which is associated to the eigenvalue 
exiting the unit circle (in complex plane) at $-1$, is (exponentially) 
{\em localized} at the center of the rotobreather and then, a new 
(linearly stable) rotobreather with frequency $\omega_b/2$ exists past 
the bifurcation. This new rotobreather can be easy and {\em safely} 
obtained by slightly perturbing the unstable rotobreather along the 
direction of the destabilizing eigenvector. In other words, although 
one cannot continue the localized solution in a bifurcation, local 
bifurcation analysis helps to throw a bridge over the bifurcation, 
so arriving safely to the new localized solution at the other side.

Continuously varying the external frequency $\omega$, further period doubling
bifurcations are often found leading to a chaotic solution. In order to
characterize unambiguosly this solution as chaotic, we have computed its 
{\em Lyapunov spectrum} $\{\lambda_i\}$, which is shown in 
figure~\ref{Spectrum}. 
\begin{figure}
\vbox to 4cm{\vfill\centerline{\fbox{Here is the figure}}\vfill}
\caption[]{Lyapunov spectrum of a chaotic rotobreather in the JJ ladder.}
\label{Spectrum} 
\end{figure}
There is only one positive Lyapunov exponent, $\lambda_1 = 0.049$bits/s. 
As we are dealing here with a continuous time dynamical system, a null 
exponent is also present. The rest of the spectrum is negative. 
Thus, there is only one expanding direction (degree of freedom) 
in phase space. The estimated Lyapunov dimension, $D_L$, defined\cite{Ott} as
\begin{equation}
D_L = j + \frac{1}{|\lambda_{j+1}|}\sum_{i=1}^{j}\lambda_i
\label{Lyap}
\end{equation}
with $j$ such that $\sum_{i=1}^{j}\lambda_i>0$ and 
$\sum_{i=1}^{j+1}\lambda_i<0$ (exponents are ordered in decreasing order), 
is $D_L= 4.7$. 

A look at the profile (at different times) of the Lyapunov vector 
associated with the positive Lyapunov exponent reveals that it is 
strongly localized in space. As the period doubling bifurcations leading to 
the chaotic solution are driven by exponentially localized 
eigenvectors of the Floquet matrix, it is not surprising that this 
chaotic solution is exponentially localized. In figure~\ref{strange} 
we show the Poincar\'e (stroboscopic, with period $2 T_b$) section 
of the central rotor trajectory $\phi_0(t) \pmod{2\pi}$ of the 
intrinsically localized chaotic solution for parameter values 
$J_x = 0.05$, $J_y=0.5$, $\epsilon_x=0.03$, $\epsilon_y=0.01$, 
$\omega=1.623$ and $I_{ac}=0.72$. As shown also in figure~\ref{strange}, the 
trajectories $\phi_i(t)$ for $|i|>0$ are noisy (or chaotically perturbed) 
oscillations. As a rough measure of "noise intensity", we adopt the radius 
$r_{i}$ of the smallest circle containing the Poincar\'e section of the 
$i$th oscillator. This quantity decreases exponentially $r_{i}\simeq C 
\exp(-|i|/\xi)$ $(\xi\sim 1.13)$, as evidenced in figure~\ref{strange}.
\begin{figure}
\vbox to 7cm{\vfill\centerline{\fbox{Here is the figure}}\vfill}
\caption[]{Poincar\'e (stroboscopic) sections of the trajectories $\phi_i (t)$
at times $t+n2 T_b$. The pictures show in quite different scales the planes
$\dot \phi$ (in radians per unit time) vs. $\phi$ (in radians) for the rotor 
(a), its fifth neighbour (b) and the ninth one (c). In (d), the logarithm of the 
``noise amplitude'' plotted vs. the neighboring index shows the exponentially 
localized character of chaos.}
\label{strange}
\end{figure}

Vaguely speaking, one could say that the uniformly oscillating solution is 
robust enough to exponentially damp out the penetration of the chaotic 
perturbation produced by the central rotor; equivalently, one could say 
that the uniformly oscillating solution posses {\em finite coherence length} 
$\xi$, so that an oscillator does not feel the effect of any sustained local 
perturbation located at distances much greater than $\xi$ (lattice units) 
from it. On intuitive basis, it is clear that finite coherence length is 
required for intrinsic localization to occur\cite{FloriaMazo}. 

Now we turn to the question on genericity, {\em i. e.} should one expect 
that these intrinsically localized chaotic solutions exist generically 
in discrete arrays of coupled nonlinear oscillators? Though arguably there 
is little 
doubt that finite coherence length is ubiquitous in discrete nonlinear 
extended systems, at least some degree of robustness of the chaotic 
trajectory in the central oscillator (not to speak of the mere possibility 
of a chaotic behaviour) is also needed. In an attempt to pave the way 
towards rigorous answers to the question, we have considered the 
perspective on intrinsic localization opened by the "anticontinuum 
limit"~\cite{Aubry94-95,MackayAubry,MarinAubry} approach, as explained below. 

Let us consider a chain of forced and damped identical {\em uncoupled} 
oscillators, and assume that there is coexistence of a chaotic attractor 
and an attracting cycle in the single oscillator phase space. Now, consider 
the cartesian product of a (central site) chaotic attractor and attracting
regular cycles in the rest of lattices sites. This set is an attractor in
the phase space of the uncoupled chain, which could plausibly be continued
when coupling is turned on.

In order to check this idea, we have chosen a chain of harmonically coupled, 
forced van der Pol oscillators:
\begin{equation}
\ddot{\phi}_i =
 -\mu(\phi_{i}^2-1)\dot{\phi_i}-\phi_i+b\cos(\omega t)
 +C(\phi_{i+1}-2\phi_i+\phi_{i-1})
\label{vanderPol}
\end{equation}
For $\mu=4.033$, $b=9.0$ and $\omega=\pi$, the single forced van der Pol 
oscillator phase space shows coexistence of two strange attractors and a 
periodic cycle of frequency $\omega/3$ (see \cite{VdP}). We have numerically 
continued the solution of the uncoupled chain in which the central oscillator 
follows a chaotic trajectory in one of the strange attractors, while the 
rest of the oscillators follow the periodic cycle, for non-zero values of the 
coupling constant $C$, up to values of the order of $0.5 \times 10^{-3}$, which 
are small but significantly different from zero. 

The continuation from the uncoupled limit provides a systematic way of 
obtaining intrinsically localized chaotic solutions, provided the coexistence 
of strange and periodic attractors for a single oscillator. It may also serve, 
like in the simpler case of periodic discrete breathers, as a basis for the 
construction of a proof of existence which we see as a difficult problem.
Indeed, Mackay \cite{Plykin} already mentioned this approach for the 
case of the Plykin attractor, where continuation is ensured due to 
uniform hyperbolicity; unfortunately, as usual in chaos theory, strong 
conditions which simplify mathematical proofs do not seem to fit easily 
into realistic physical models.

The examples we have shown here concern systems of forced and damped 
oscillators, and one may wonder about hamiltonian arrays of oscillators. 
Though we do not have a definite answer on the existence of intrinsic 
localized chaos in discrete nonlinear Hamiltonian extended systems, it seems 
plausible that the typical "broad band" structure of the power spectrum of 
chaotic trajectories would imply a violation of the condition of non-resonance 
with the phonons \cite{MackayAubry}. In the extent that this condition plays 
an essential role in the proof of existence of hamiltonian discrete breathers, 
we think that the answer is negative. However, chaotic breathers in discrete 
Hamiltonian arrays easily appear as long-lived transient solutions. An
observation of erratically moving transient chaotic breathers in hamiltonian
Fermi-Pasta-Ulam chains has been recently reported\cite{Dauxois}. After
completion of this work, we became aware of the numerical observation of  
chaotic rotobreathers by Bonart and Page\cite{Page} in a 1d driven damped 
lattice of dipoles.

\stars
We acknowledge to S. Aubry, C. Baesens, R.S. Mackay and J.L. Mar\'{\i}n 
for many useful discussions, 
P. Grassberger for his illuminating criticisms and J. Page for sending us a 
draft of his work, prior to publication. This work has been financially 
supported by DGES through project PB95-0797. One of us (JJM) acknowledges 
a Fulbright-MEC fellowship.

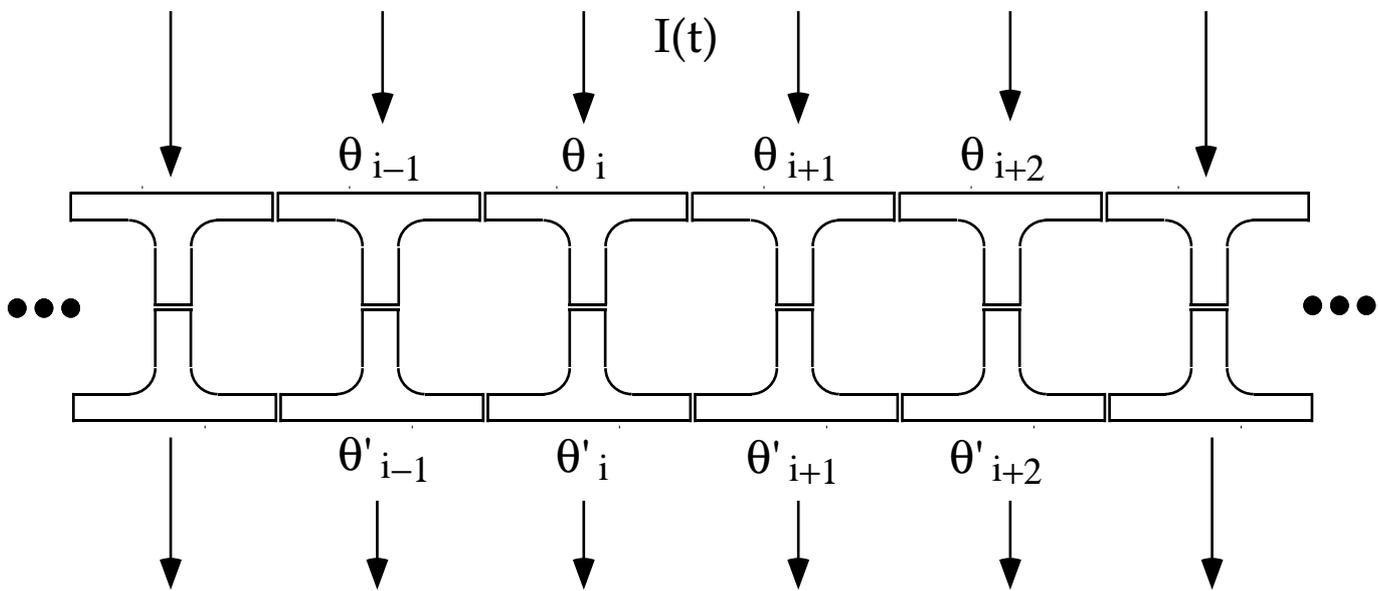
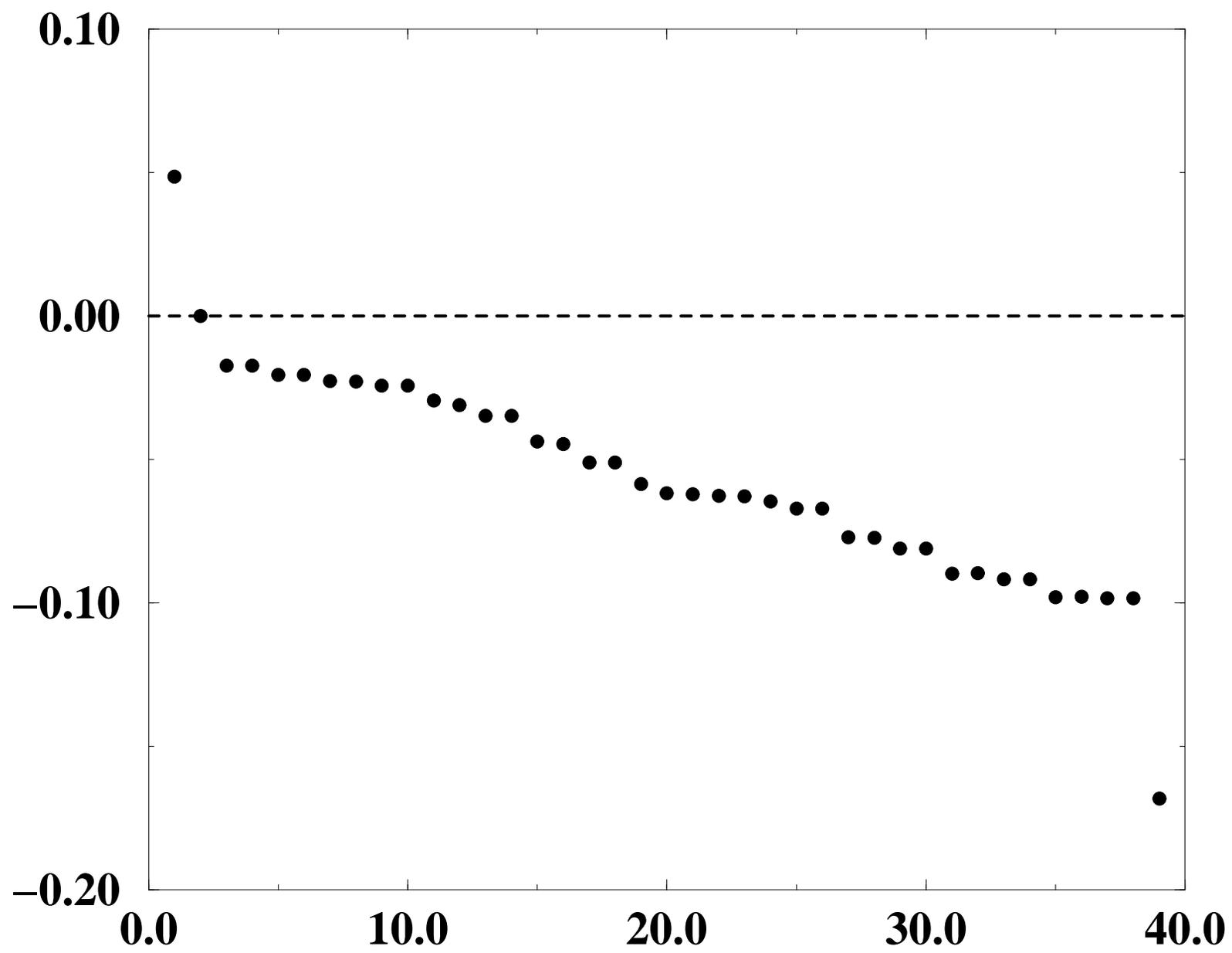
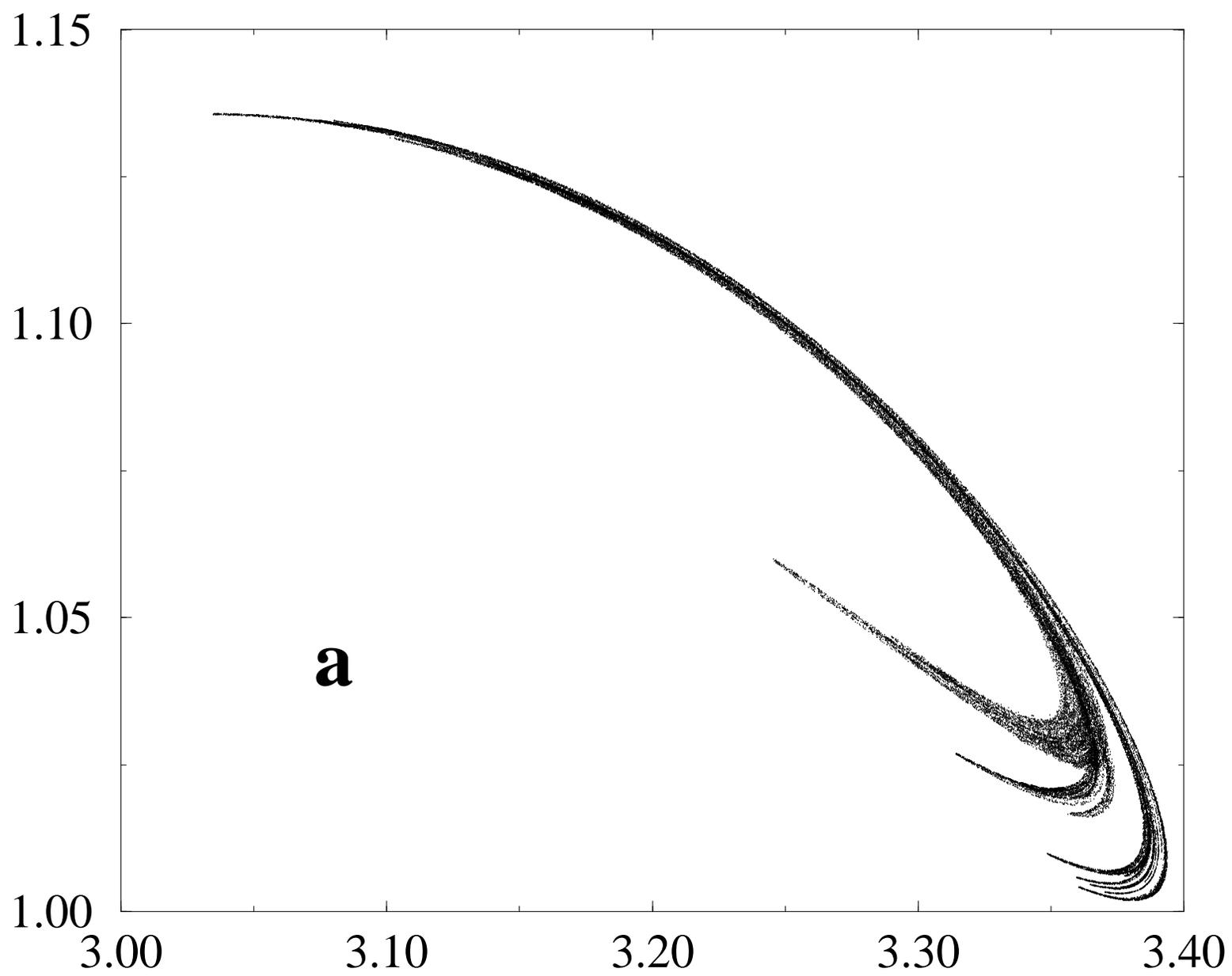
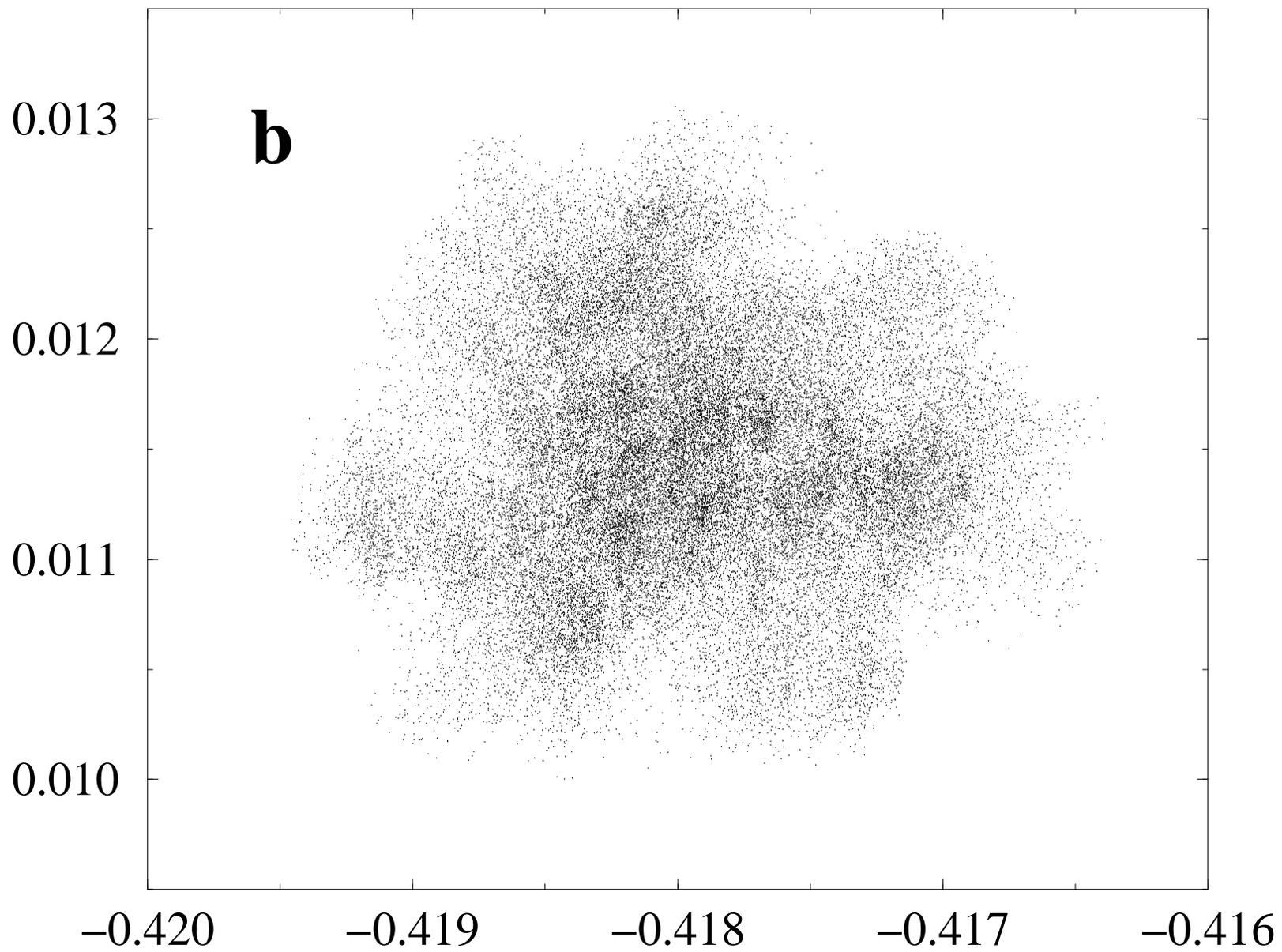
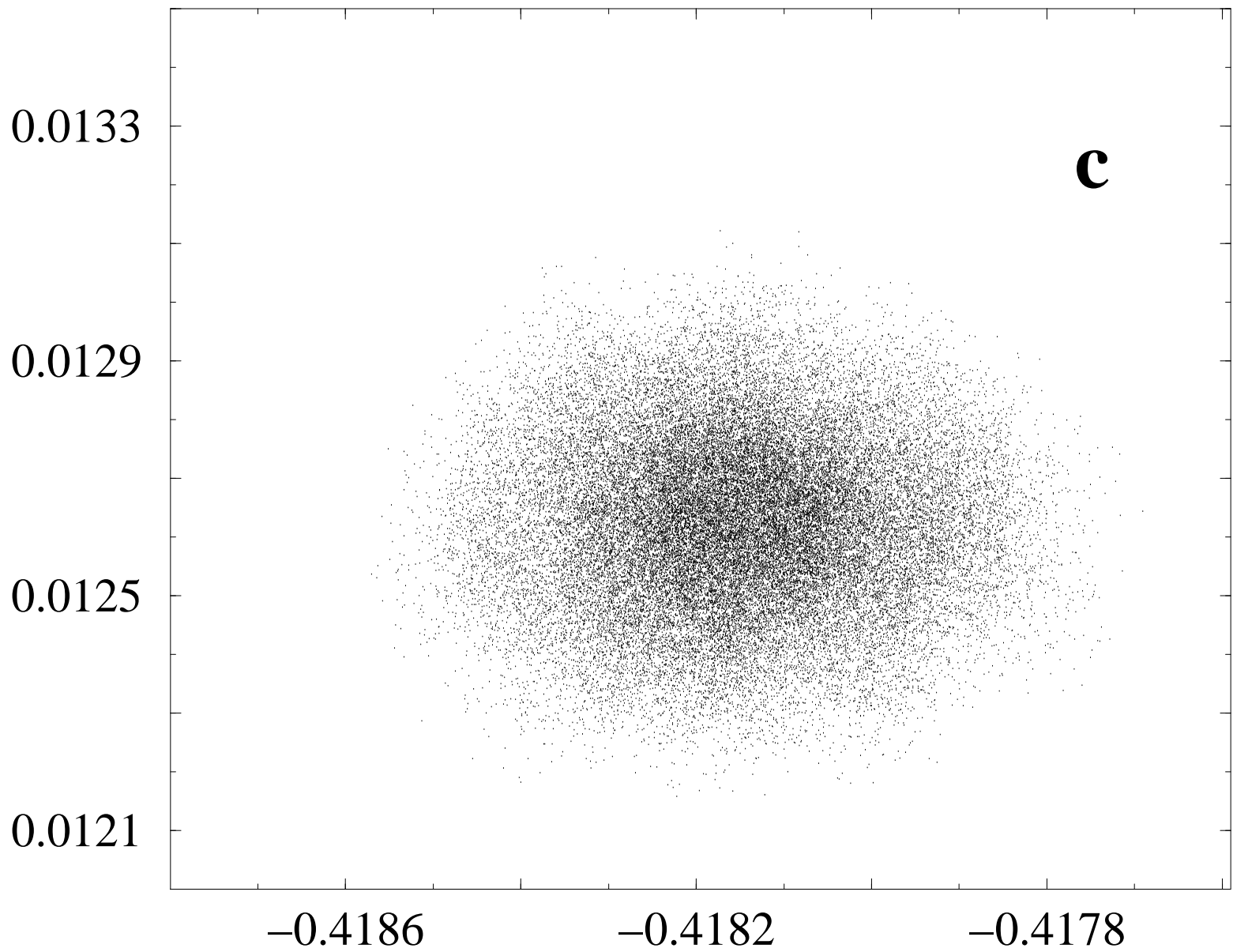
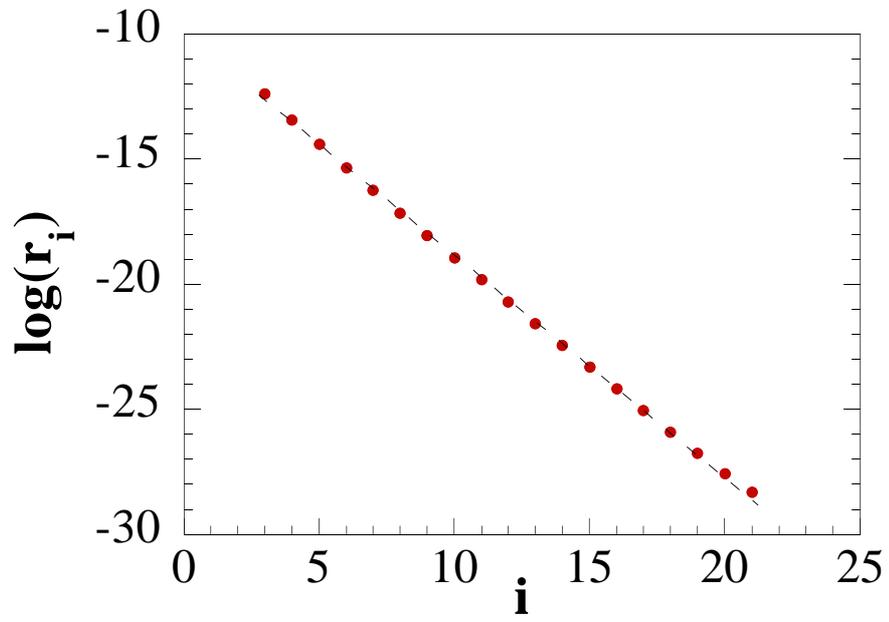
\end{document}